\documentclass[prl,twocolumn,superscriptaddress,amsmath,amssymb]{revtex4-1}
\usepackage{graphicx}
\usepackage{color}

\begin{document}
\title{Easier sieving through narrower pores: fluctuations and barrier crossing in flow-driven polymer translocation}

\author{\firstname{R.} \surname{Ledesma-Aguilar}}%
\email{r.ledesmaaguilar1@physics.ox.ac.uk}
\affiliation{Rudolf Peierls Centre for Theoretical Physics, University of Oxford, 1 Keble Road, Oxford OX1 3NP, United Kingdom}
\author{\firstname{T.} \surname{Sakaue}}
\affiliation{Department of Physics, Kyushu University 33, Fukuoka 812-8581, Japan}
\affiliation{PRESTO, JST, 4-1-8 Honcho, Kawaguchi, Saitama 332-0012, Japan}
\author{\firstname{J. M.} \surname{Yeomans}}
\affiliation{Rudolf Peierls Centre for Theoretical Physics, University of Oxford, 1 Keble Road, Oxford OX1 3NP, United Kingdom}
\date{\today}

\begin{abstract}
We show that the injection of polymer chains into nanochannels becomes easier as the channel becomes narrower. This counter intuitive result arises because of a decrease in the diffusive time scale of the chains with increasing confinement. The results are obtained by extending the de Gennes blob model of confined polymers, and confirmed by hybrid molecular dynamics - lattice-Boltzmann simulations.
\end{abstract}

\maketitle


The dynamics of confined polymer chains in solution is of great importance  to many biological processes. These range from the 
permeation of macromolecules across the nuclear pore complex~\cite{Talcott-TrendsCellBio-1999}, to their packaging and ejection from viral capsids~\cite{Kindt-PNAS-2001,Smith-Nature-2001}, and their motion in the crowded cellular environment~\cite{Ellis-TrendsBiochemSci-2001}.  

A problem that has captured the attention of the physics community in recent years 
is the translocation of polymer chains through narrow pores~\cite{Meller-JPhysCondensMatt-2003}, an ubiquitous process in 
cellular biology that has enormous potential in developing controlled applications such as 
the separation of macromolecules in designed microfluidic devices~\cite{vanRijn-Nanotechnology-1998} or the selective recognition and sequencing of DNA \cite{Branton-NatBiotechnol-2008}.  
While much effort has been devoted to characterise the transient passage through very short pores~\cite{Graham-AnnuRevFluidMech-2011,Fyta-JPolyMatSciB-2011,Sakaue-PRE-2010,Milchev-JPhysCondensMatter-2011}, for long channels~\cite{Luo-JChemPhys-2011} questions  such as whether the chain translocates at all, and how this process is triggered,
have been explored to a much lower degree.    

From a theoretical perspective, it is now well accepted that the injection of chains into long channels involves 
the surpassing of a free energy barrier~\cite{Meller-JPhysCondensMatt-2003,Fyta-JPolyMatSciB-2011}, which for flow-driven 
chains is set by the competition of the driving hydrodynamic 
drag and the entropic pressure resisting the deformation of the chain as it is squeezed into the constriction.  
Within the de Gennes blob model~\cite{deGennesScaling},  Sakaue {\it et al.} ~\cite{Sakaue-EPL-2005} showed that the 
barrier is overcome above a critical flow rate, which, remarkably, is independent of the length of the chain 
and the geometrical features of the nanochannel.

In this Letter we highlight the importance of hitherto unknown scaling aspects of 
the barrier-crossing dynamics by exploring the 
Brownian motion of the chain within the channel.   We extend the de Gennes blob model 
beyond the current static description,  
and include the effect of the attempt frequency of the chain to cross the free energy  barrier.  
Surprisingly, we find that the threshold flow rate decreases with decreasing diameter of the nanochannel, making translocation more probable across narrower constrictions. This 
unexpected result is due to a decrease in the diffusion timescale of the blob-like chains as the diameter of the channel is decreased, therefore increasing the attempt frequency of the chain to cross the free energy barrier. To confirm this prediction, we have carried out hybrid molecular dynamics - lattice-Boltzmann simulations~\cite{Ahlrichs-JChemPhys-1999,Usta-JChemPhys-2005}, finding a very good agreement with our theory.

This novel and counter intuitive feature of the translocation process has relevance in nano- and biotechnological applications.  While 
we focus on the flow-driven translocation of the chains, the enhancement of translocation in narrower pores is independent of the driving
force, and is thus expected to hold for other systems involving confined polymer chains in solution, for example, for electrophoretic~\cite{Sakaue-EPJE-2006}
and diffusiophoretic~\cite{Palacci-PhysRevLett-2010} driving.

{\it Blob model for polymer translocation.--} Exact analytic or fully resolved numerical approaches to the dynamics of a confined polymer chain in solution are, in general, not 
feasible. The chain itself contains a large number of atoms and the complexity of the problem is increased even further by the need 
to include the  hydrodynamic interactions mediated by the fluid.  

To make progress, coarse graining is needed, where the number of degrees of freedom is drastically reduced in a way that preserves the
relevant physics in a given problem. Our approach here is to extend the de Gennes blob model to study the
flow-driven translocation of the chains.   

\begin{figure}[t!]
\includegraphics[width=0.45\textwidth]{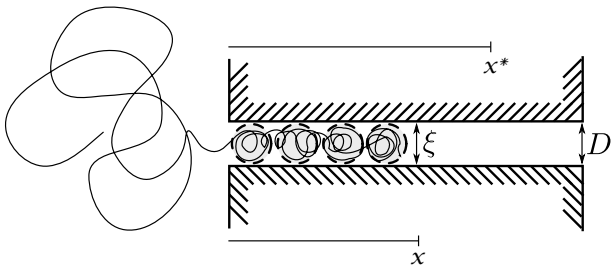}
\caption{(Color online) Blob picture for the penetration of a linear polymer chain into a nanochannel driven by a convergent fluid flow.  When pushed
into a nanochannel of diameter $D$, the polymer chain deforms into an elongated configuration composed of
a series of space-filling blobs of size $\xi \simeq D$.  As the penetration length, $x$, increases, the chain can reach
the position of the free energy barrier, $x^*$, where hydrodynamic driving takes over the resisting entropic pressure, 
thus triggering the translocation of the polymer.    
\label{fig:Diagram0}}
\end{figure}

We start by considering a linear polymer composed of $N$ monomers joined by links of size $b$, which lies constrained 
by a channel of diameter $D > b$.  We focus in the limit of small pores, where the chain is deformed from its equilibrium radius 
of gyration, $R\simeq b N^{3/5}\gg D$, adopting an elongated shape of length $L$. 
The blob model assumes that the chain can be regarded as a sequence of $M$  
blobs, each of size $\xi$ (see Fig.~\ref{fig:Diagram0}).  It then follows that the extension of the chain inside the channel is
$L\simeq \xi M.$
Within each blob, the effect of confinement is assumed to be unimportant.  Therefore, the size of the blob 
scales with the number of monomers in a blob, $P$, following the Flory prediction, 
$\xi \simeq b P^{3/5}.$
Using the conservation of the number of monomers, $N=MP$, one can obtain a scaling law for the length of the chain:
\begin{equation}
L \simeq \xi \left(\frac{b}{\xi}\right)^{5/3}N.
\label{eq:Lblob1}
\end{equation}
This result is consistent with the Flory free energy of the confined chain,
 \begin{equation}
\frac{F}{k_{\rm B} T} \simeq \frac{L^2}{b^2 N}+\frac{N^2 b^3}{LD^2}, 
\label{eq:ConfinedFloryEnergy}
\end{equation}
where $T$ is the temperature and $k_{\rm B}$ is Boltzmann's constant.
Eq.~(\ref{eq:ConfinedFloryEnergy}) describes the competition between the entropic elasticity of the chain conformation, 
given by the first term, and  the internal collisions that give rise to excluded volume effects, represented by the second term.  
The length of the chain is set by the balance between both contributions, giving the scaling  
$L\simeq D \left(\frac{b}{D}\right)^{5/3}N,$
which matches the blob model result, Eq.~(\ref{eq:Lblob1}), for {\it space-filling chains, i.e.} where $\xi \simeq D$.

Turning our attention to the translocation process, we consider a polymer which is
brought to the entrance of the constriction by an underlying fluid flow and partially enters it up to a distance $x$, as shown in Fig.~\ref{fig:Diagram0}. 
If the free portion of the chain is not significantly deformed by the fluid pressure, the only competing mechanisms governing
the motion of the polymer are the entropic cost of confinement, which tends to increase the free energy of the chain, $\Delta F$, 
and the drag exerted by the driving fluid, which tends to reduce it.  Assuming an entropic contribution of $k_{\rm B}T$ 
per confined blob, and that the fluid drag acting on each blob scales as $\eta J \xi/D^2$, it follows that
$\Delta F$ depends on $x$, according to
$\Delta F(x) \simeq  k_{\rm B} T \left(\frac{x}{D}\right) - \frac{\eta J}{2} \left(\frac{x}{D}\right)^{2}$~\cite{Sakaue-EPL-2005};
here $J$ is the flow rate (measured in units of volume per unit time) and $\eta$ is the dynamic viscosity of the fluid.   
It is clear from the interplay between the two terms in $\Delta F$ that there is a free energy barrier, $\Delta F^* \simeq \left(k_{\rm B} T\right)^2/(\eta J),$ which is overcome once the chain has reached a critical penetration length 
$x^* \simeq \frac{k_{\rm B} T}{\eta J}D.$

Sakaue {\it et al.}~\cite{Sakaue-EPL-2005} considered the case where the strength of the 
barrier is of the order of thermal fluctuations, {\it i.e.,} 
$\Delta F^* \simeq k_{\rm B}T$.  This leads to $x^*\sim D$, 
showing that the translocation is triggered when a single blob has moved into the channel.   It follows 
from this assumption that the barrier is surmounted above a critical flow rate 
\begin{equation}
J_c (x^*=D) \simeq \frac{k_{\rm B} T}{\eta}.
\label{eq:CriticalFlux}
\end{equation}
The striking feature of this leading-order scaling
prediction is that $J_c$ depends on neither $N$ nor $D$,   as long as $L\gg x^*$~\cite{Markesteijn-SoftMatter-2009,Ledesma-SoftMatter-2012}.

However, this first approximation disregards the fluctuating motion of the chain in the constriction, 
which is responsible for a crucial dependence of the translocation on the diameter of the nanochannel. 
Even if $J < J_c$,  the chain can still overcome the barrier due to thermal fluctuations.  
This is easily seen by considering the probability of chain translocation,
$P(J) = \kappa \tau_{\rm m} \leq 1,$
where $\tau_{\rm m}$ is the observation time and $\kappa = \kappa_{D}\exp\left\{-{\Delta F^*}/{k_{\rm B} T}\right\}$ 
is the translocation rate over the barrier.
The transition rate contains the exponential factor indicative of the free energy barrier, and the  
{\it attempt frequency}, $\kappa_{D}$, which reflects the fluctuation of the position of the chain 
within the channel.  This additional contribution to $P(J)$ is particularly important close to the penetration threshold, where the forces acting on the polymer effectively balance each other out leaving the blobs in the chain free to undergo a Brownian motion over length scales comparable to their own size.  The mean square displacement of the blobs evolves in time, $t$, according to $\langle x \rangle ^2 \sim D_{\rm blob} t$, 
where the diffusion coefficient obeys Einstein's relation, 
$D_{\rm blob}\sim k_{\rm B} T /(\eta \xi).$  In this limit the attempt frequency follows by taking
the reciprocal of the timescale of self-diffusion of one blob, $\kappa_D\simeq D_{\rm blob}/\xi^2$. 
Recalling that for a space-filling chain $\xi \simeq D$, we get
\begin{equation}
\kappa_D \simeq \frac{k_{\rm B} T}{\eta D^3}.
\label{eq:kappaD}
\end{equation}

To demonstrate the relevance of this result let us define a {threshold current}, $J^*<J_c$,  {\it via} the relation
$P( J^*) = P^*,$
where $P^*<1$ is an arbitrary value of the probability of translocation.  After substitution of $\kappa_D$ we obtain 
\begin{equation}
J^*\simeq  \frac{J_c}{\ln\left(\frac{X}{D}\right)},
\label{eq:ThresholdFlux}
\end{equation} 
where $X^3=k_{\rm B}T\tau_{\rm m}/(\eta P^*)$. We thus find a
logarithmic correction to the leading-order scaling prediction  
which gives the dependence of the translocation rate on the observation time, {\it e.g.,} the experimental timescale, as well as on the  channel size, $D$.  
According to Eq.~(\ref{eq:ThresholdFlux}), $J^*$ is only weakly sensitive to $D$ when the
observation time is long enough. In this limit, the simple scaling
structure of Eq.~(\ref{eq:CriticalFlux}) is recovered. However, if the
observation time becomes comparable to the blob time scale $\sim \kappa_D^{-1}$, 
then $J^*$ increases sharply with $D$.


{\it Molecular dynamics/lattice-Boltzmann simulations.--} In order to confirm the predicted dependence of the threshold flow rate on the 
nanochannel thickness, we have carried out numerical simulations of confined chains in narrow channels, in equilibrium and 
in flow-driven configurations, using a hybrid molecular dynamics - lattice-Boltzmann algorithm.  

We consider a bead-spring model for the polymer.  The chain is coarse-grained into $N$ solid beads of radius $a$ joined by $N-1$ massless springs.  The dynamics of the chain is given in inertial form,
\begin{equation} 
m\ddot{\bf r}_i= -\sum_{j\neq i} {\boldsymbol \nabla}_{ij} U + {\bf F}_i,
\label{eq:MD}
\end{equation}
where ${\bf r}_i$ is the position vector of the $i$-th bead, $m$ is its mass and the double dot symbol stands for double differentiation with respect to time.  The first term in the right-hand side contains the elastic force between adjacent  beads in the chain,  which are modelled using a Hookean potential with elastic constant $k$ and rest length $b$, and the excluded volume interactions between all beads in the 
chain which follow from a DLVO potential, 
$U_{\rm DLVO}=U_0\exp\left\{-\kappa_{\rm DH} (r_{ij}-2a)\right\}/({r_{ij}-2a}),$
where $r_{ij}$ is the distance between beads $i$ and $j$, $\kappa_{\rm DH}$ is the Debye-H\"uckel screening length and $U_0$ is an amplitude.  The same 
potential is used to model repulsive interactions between the beads and solid walls. 
  
The remaining term, ${\bf F}_i={\bf F}_i^{\rm r} + {\bf F}_i^{\rm h}$, accounts for the random and viscous forces, 
${\bf F}_i^{\rm r}$ and $ {\bf F}_i^{\rm h}$, that originate from thermal fluctuations and viscous stresses within the fluid.  
The random force, ${\bf F}_i^{\rm r}$, has zero mean and a time-correlation matrix satisfying the  
fluctuation-dissipation relation,  
$\langle {\bf F}_i^{\rm r}(t){\bf F}_i^{\rm r}(t') \rangle = 2k_{\rm B} T \zeta \delta(t-t')\bf 1,$
where $\zeta$ is the friction coefficient of a bead with the fluid, $\delta(t)$ is Dirac's delta function and $\bf 1$ is the 
identity matrix. 
The viscous force, ${\bf F}_i^{\rm h}$, is modelled 
using Stokes law, which compares the velocity of the bead, $\dot {\bf  r}_i$, and the velocity of the fluid at the position
of the bead, ${\bf v}({\bf r}_i)$; 
${\bf F}_i^{\rm h} = -\zeta (\dot{\bf r}_i - {\bf v}({\bf r}_i)).$ 
Here $\zeta=6\pi\eta r_{\rm h}$ is the Stokes drag coefficient, where $r_{\rm h}$ is the hydrodynamic radius of the beads.   
 
The dynamics of the solvent is modelled using the fluctuating lattice-Boltzmann algorithm for a Newtonian 
fluid~\cite{Ahlrichs-JChemPhys-1999,Usta-JChemPhys-2005}, which we solve simultaneously to Eq.~(\ref{eq:MD}). 
This allows us to compute the friction force, ${\bf F}_i^{\rm h}$, on all beads, which is
exerted back onto the fluid to ensure global momentum conservation.   The bead mobility $\sim1/6\pi\eta a$
is recovered by including a lattice correction to $r_{\rm h}$.  


{\it Confined chains in equilibrium.-} We first compare the equilibrium properties of polymers under full confinement to the blob-model
prediction.  We consider a chain in a periodic simulation box.  In lattice units, the dimensions of the box are $240\times D\times D$ along the $x$, $y$ and $z$ directions, respectively.  Also in lattice units, chain parameters are fixed to $N=128$, $a=0.25$, $b=1.0,$ $k=30k_{\rm B}T,$ and $m=17 (4a^3/3)\rho$.  DLVO parameters are fixed to $U_0=k_{\rm B}T$ and $\kappa_{\rm DH} = 80$, which ensures 
a repulsive force between beads comparable to $k_BT$ at distances between beads $\simeq 0.5a$, preventing chain crossings.  The viscosity, density and temperature of the fluid bath are chosen as $\eta=6$, $\rho=36$ and $k_{\rm B}T=0.1$, while we take the hydrodynamic radius as $r_{\rm h}=0.32$.  

The initial configuration is prepared by placing a 128-bead chain ($R\approx 8$) in a wide channel ($D=16$); we subsequently
apply a small forcing along the negative $x-$, $y-$ and $z-$directions and allow the chain to relax to a `compressed' state which
is then used as the initial condition in our simulations, reducing the likelihood of trapping in metastable states. 
Equilibration is then carried out for several values of $D$ and $N$ over $1.25\times10^4$ timesteps, during which we record the length of the chain, $L$, by measuring the difference between the maximum and minimum $x-$coordinates accessed by the polymer at a given time.  Data is averaged over fifty
realisations of the noise. 

\begin{figure}
\includegraphics[width=0.23\textwidth]{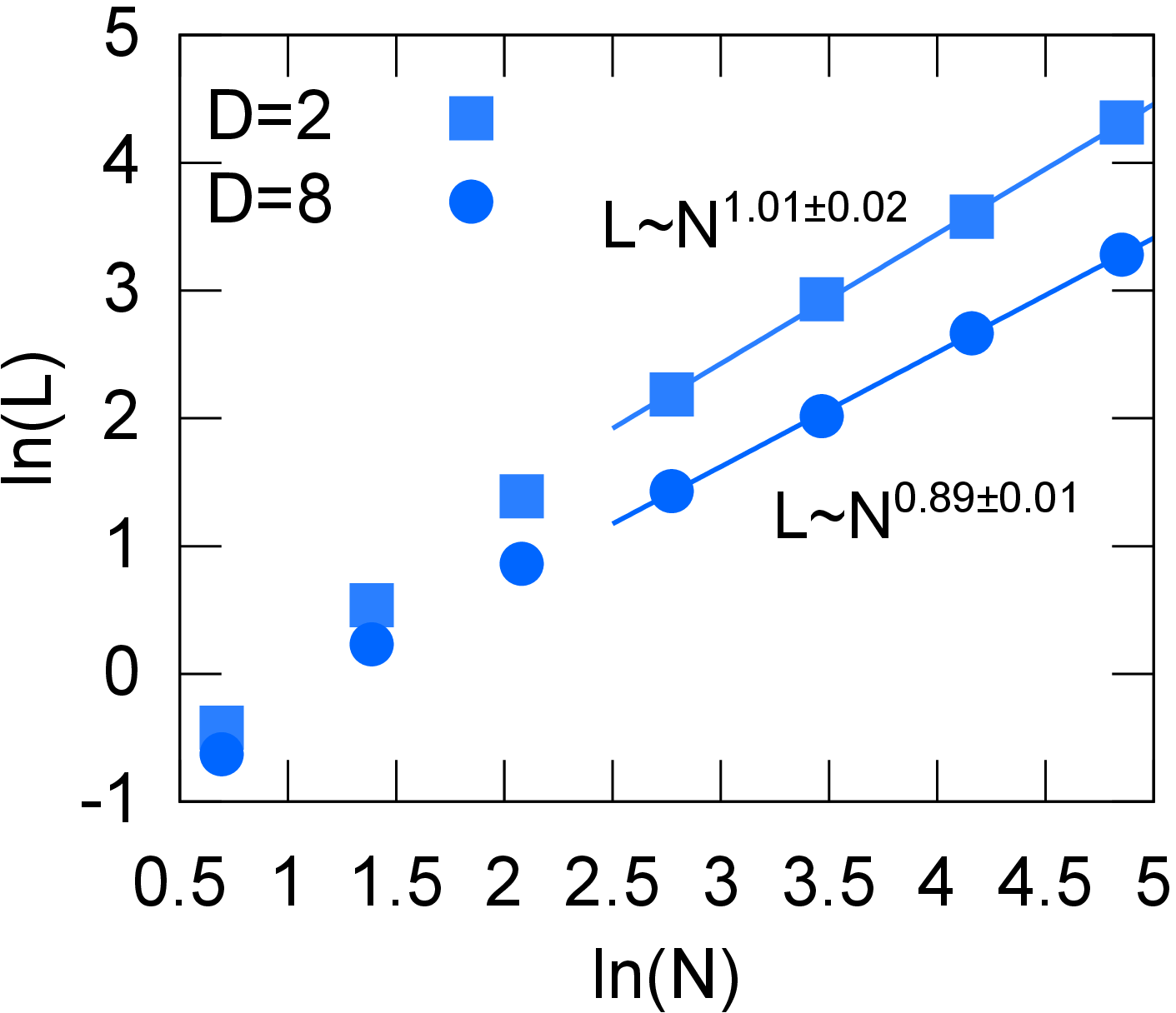}~~~~\includegraphics[width=0.23\textwidth]{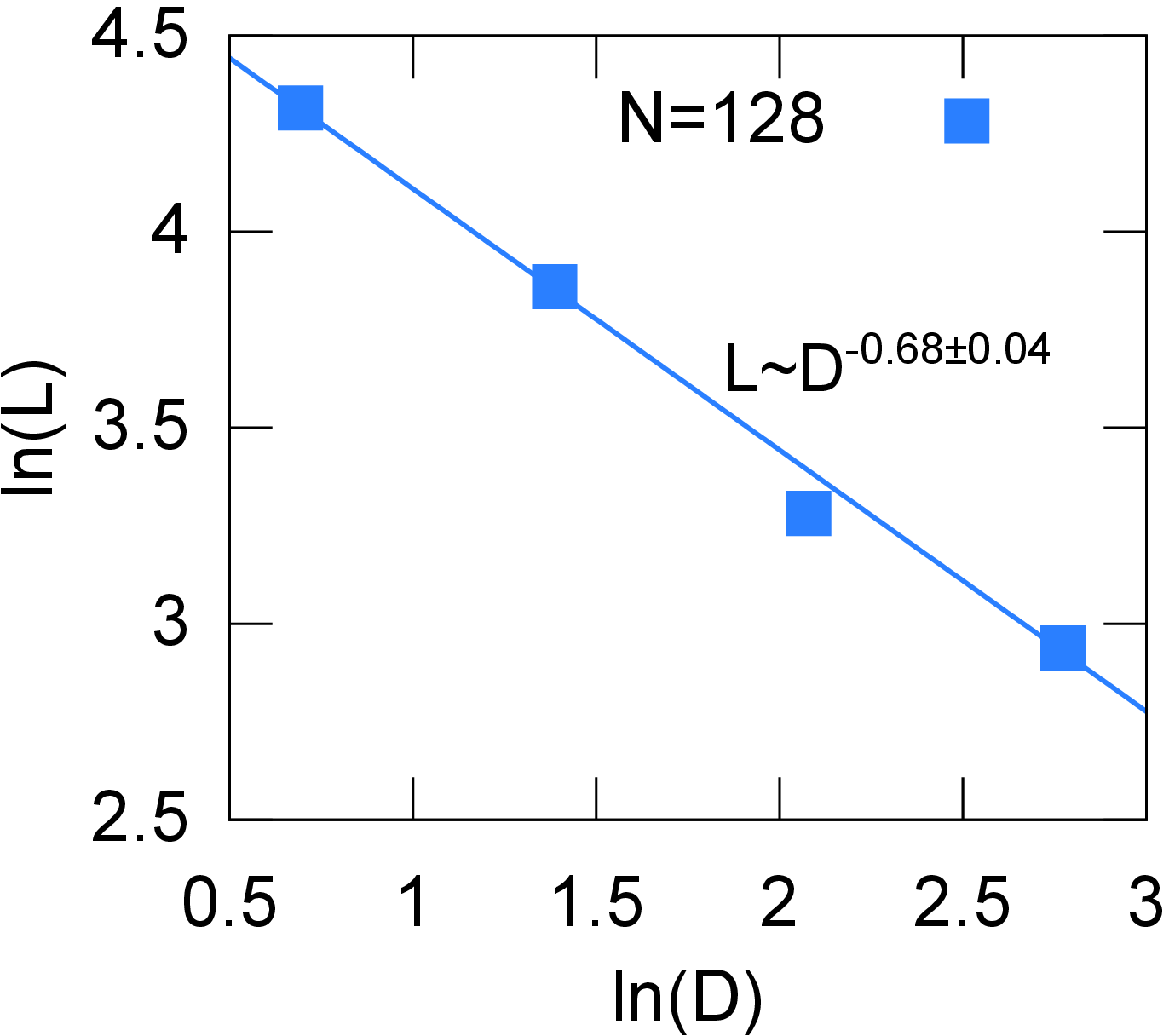}\\
(a)~~~~~~~~~~~~~~~~~~~~~~~~~~~~~~~~~~(b)
\caption{(Color online) (a) Length of a confined chain $L$, as a function of the number of beads in the chain, $N$, for two channels of different width, $D$.  For large $N$ and small $D$, simulation results (symbols) approach the blob-model prediction, characterised by a scaling exponent $\nu = 1$.   (b) Length of a chain as a function of the channel width. Symbols correpond
to simulation results showing a good agreement with the exponent of $-2/3$ predicted by the blob model.  
Solid lines in both panels are a linear fit to the data. Error bars are smaller than the symbol size.
\label{fig:L.vs.D}}
\end{figure} 
Following previous numerical studies~\cite{Kremer-JChemPhys-1984,Milchev-MacromolTheorySimul-1994,Lendrejack-PRL-2003,Chen-PRE-2004,Wang-Macromolecules-2011}, we expect a scaling behaviour $L\sim N^\nu$, where the $\nu$-exponent should cross over from the 3D value $\nu \approx 0.59$
to the 1D value $\nu=1$ as the confinement becomes stronger.  Our numerical simulations confirm the scaling prediction for 
the confined regime,  
as can be seen in Fig.~\ref{fig:L.vs.D}(a),
where the upper curve, corresponding to 
$D=2$, already shows a linear increase for moderate values of $N$.   For $D=8$ we observe a slightly smaller
exponent of $\nu = 0.89\pm0.01$;  such proximity to the fully developed 1D value, even for weakly confined chains 
($D\sim R \approx 8 $),  indicates that the crossover occurs rather quickly, both as a function of $D$ and $N$.  
This assertion is further supported by looking at the dependence of the length of the chain on the channel width for relatively long chains ($N=128$),which shows the expected power-law behaviour with an exponent $ -0.68 \pm 0.04 $, very close to the blob model prediction,
 $L\sim D^{-2/3}$ (see Fig.~\ref{fig:L.vs.D}(b)).
   
{\it Forced translocation.--} To study the translocation process, we equilibrate a 128-bead chain for 
$1\times10^4$ timesteps on the {\it cis} side of a two-compartment duct of dimensions 
$240\times14\times10$ separated by a channel of dimensions $24\times D\times D$.  We disregard 
the process of finding the pore, and carry out the equilibration stage while keeping the first bead of the polymer 
tethered to the channel entrance according to the same elastic potential as used for the beads.  
We subsequently release the chain and exert a 
constant body force, $f$, on the fluid in the $x$-direction, thus creating a convergent flow that drives the chain into 
the channel.   Simulations are carried out for a further $1\times 10^5$ timesteps, which allow for the record of 
failed (chain ejected back to the {\it cis} side) or successful (chain released to the {\it trans} side) events.  

\begin{figure}
\includegraphics[width=0.22\textwidth]{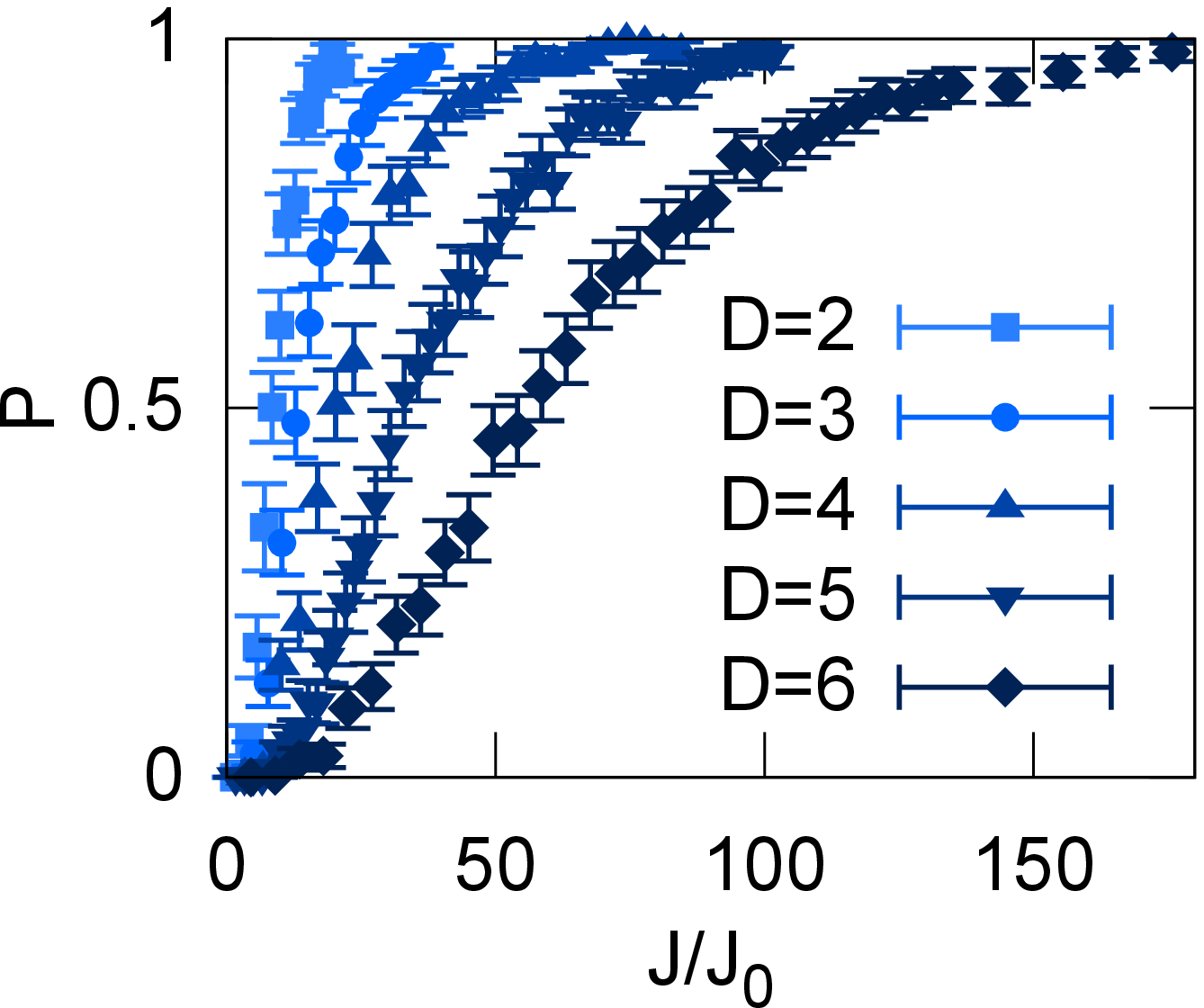}~~~~\includegraphics[width=0.22\textwidth]{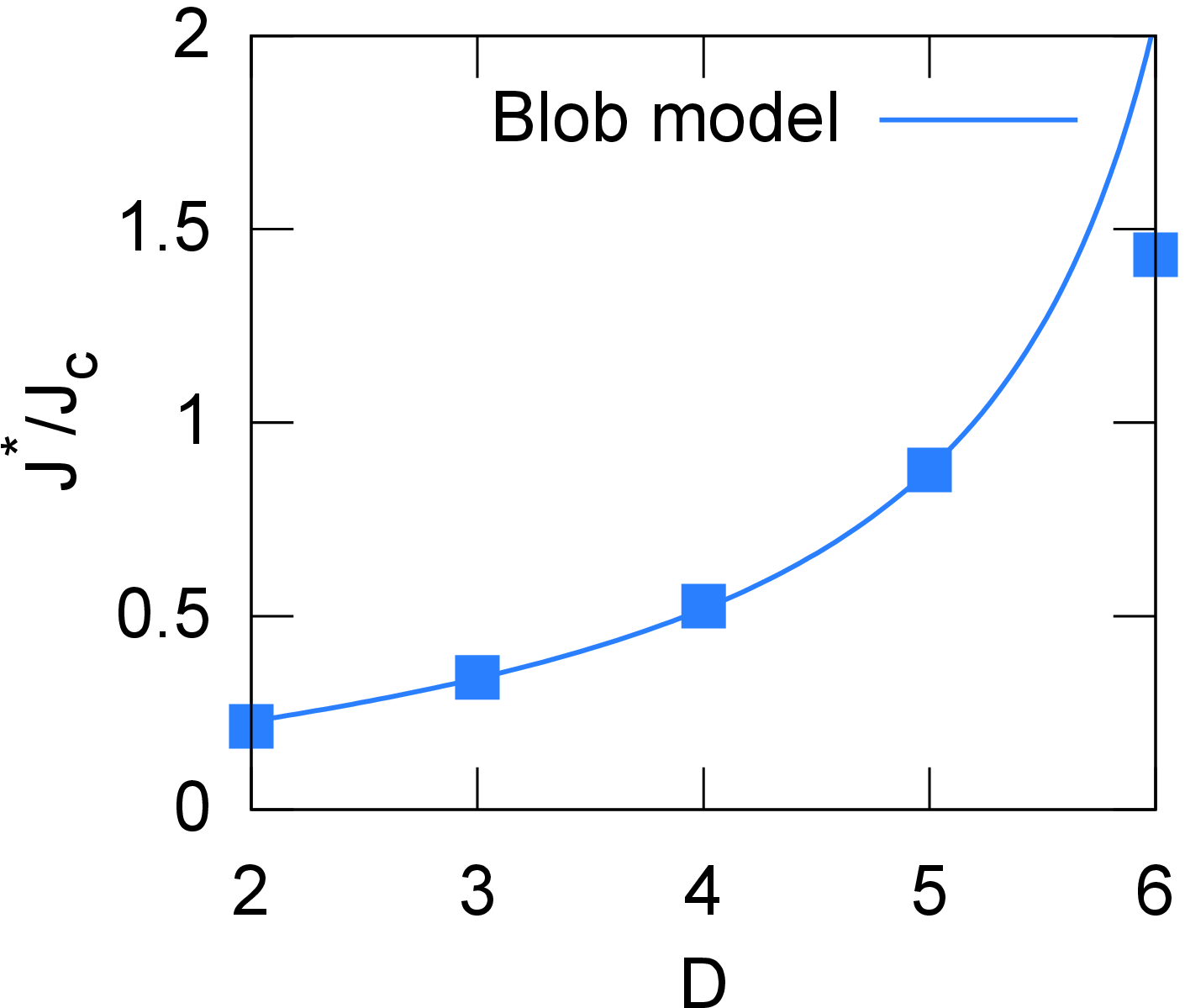}\\~
(a)~~~~~~~~~~~~~~~~~~~~~~~~~~~~~~~~~~(b)
\caption{(Color online) (a) Probability of translocation as a function of the flow rate for different pore sizes.
(b) Threshold flow rate, $J^*$, as a function of the channel width.  The threshold flow rate increases with increasing
width in agreement with the blob model (see text).  Error bars are comparable to the size of the symbol.
\label{fig:j.vs.h}}
\end{figure}
Fig.~\ref{fig:j.vs.h}(a) shows the probability of translocation of the chain, $P$, as a function of the imposed flow rate, $J$, for varying $D$.  Data are in units of a reference flow rate, $J_0\sim f_0$, where the reference
body force is set to $f_0=1\times10^{-5}$ in simulation units.   Due to the increase in the drag with $J$, the probability of translocation
increases with increasing flow rate, crossing over from a non-translocation regime ($P \rightarrow 0$) to a 
full translocation regime ($P\rightarrow 1$).  

While the observed increase of $P$ as a function of $J$ holds for all channel widths considered, the 
curves in Fig.~\ref{fig:j.vs.h}(a) shift to the right as $D$ is increased. This means that the threshold flow rate needed to trigger the translocation 
of the chain is higher in wider channels.  In order to quantify this behaviour, we define the threshold flow rate following the criterion $P(J^*)=P^*=0.5$, and measure the corresponding values of $J^*$, which, as shown in Fig.~\ref{fig:j.vs.h}(b), 
increases monotonically with $D$.    In order to compare with the blob-model prediction, we carry out a fit of the simulation data to the function $J^*/J_c = A/\ln(X/D),$ obtaining 
$A=0.28\pm0.01$ and $X=6.88\pm0.08$.  

From these fitting parameters and using Eq.~(\ref{eq:ThresholdFlux}) we obtain $\tau_{\rm m} \approx 10^4$, which  
is consistent with the order of magnitude of the 
passage time of the chain through the nanochannel ($\approx 5\times 10^4$ timesteps). 
The predicted increase of the threshold flow rate with the channel width is thus confirmed by the numerical 
results, as shown in Fig.~\ref{fig:j.vs.h}(b).  As expected, the data start to deviate from the prediction for $D=6\lesssim R$, 
where the chain begins to cross over to the unconfined regime, as found in the equilibrium simulations described above. 
It is thus pleasing that two very different approaches to coarse-graining, the blob and bead-spring models, agree for both
the statics and dynamics presented here. 


This newly reported and unexpected dependence of the threshold flow rate for polymer translocation on the size of the 
nanochannel originates from the space-filling character of the chains and their Brownian motion within the constriction. Given that the blobs composing 
the polymer match the width of the channel, the attempt frequency associated with the self-diffusion of the chain in the 
channel decreases with increasing pore width.  This is both because larger blobs self-diffuse over a larger distance, and because the viscous resistance to their motion scales with their size.  It is counter intuitive that for fully confined polymers the probability of translocation is reduced as the width of the channel is increased, leading to lower threshold flow rates in narrower channels.  

A feasible experimental setup to test our prediction, and to thereafter correctly interpret and understand experimental data, follows from B\'eguin {\it et al.}~\cite{Beguin-SoftMatter-2011}, who carried out measurements 
of the permeate flux and the rejection coefficient (related to the probability of translocation) for the injection of hydrosoluble polymers through nanopores etched in polycarbonate membranes.  
This corresponds to a blob size comparable to the size of the monomers, therefore making a direct comparison to experiments possible.   A situation of great practical importance is the confinement 
of single DNA chains in narrow channels, where the persistence length, $l_{\rm p}$, is larger than the diameter of the confining channel, $D$.  In the scaling description, the polymer length scale analogous
to the blob size is the deflection length, $l_{\rm c} \simeq l_{\rm p}^{1/3}D^{2/3,}$ which, like the blob size, is an increasing function of $D$. We therefore expect that the same qualitative results for the attempt frequency hold in 
this situation. For larger pores (in the microfluidic regime), where the persistence length of the chain is smaller than the size of the channel, the polymer behaves as a flexible chain, and is therefore also 
described by our theory. 

The threshold flow rate is a quantity of interest both from the biological and technological perspectives.
Our findings show that a detailed description of the energy barrier crossing is still necessary, and we hope to 
inspire further research on potentially important effects arising, for example, from polymer-pore 
interactions and pore geometry.   Our present results for the size dependence of the threshold flow rate may help 
in the rational design of filtering nanofluidic devices 
for which the channel size, rather than the operation flow rate, is engineered to trigger or hinder the translocation of the chains, and 
may shed light on possible mechanisms present in biological systems to control macromolecule transport. 


\end{document}